\documentclass[letter]{ptptex}

\usepackage{graphicx}

\notypesetlogo                       

\title{
Phase Transition in Potts Model with Invisible States%
}

\author{
Ryo \textsc{Tamura}$^{1,}$\footnote{E-mail:r.tamura@issp.u-tokyo.ac.jp},
Shu \textsc{Tanaka}$^{2,}$\footnote{E-mail:shu-t@alice.math.kindai.ac.jp},
and
Naoki \textsc{Kawashima}$^{1,}$\footnote{E-mail:kawashima@issp.u-tokyo.ac.jp}%
}

\inst{
$^1$Institute for Solid State Physics, University of Tokyo,
Kashiwa 277-8581, Japan\\
$^2$Research Center for Quantum Computing,
Interdisciplinary Graduate School of Science and Engineering,
Kinki University,
Higashi-Osaka 577-8502, Japan
}

\abst{
We study phase transition in the ferromagnetic Potts model with invisible states that are added as redundant states by mean-field calculation and Monte Carlo simulation. Invisible states affect the entropy and free energy, although they do not contribute to the internal energy.
A second-order phase transition takes place at finite temperature in the standard $q$-state ferromagnetic Potts model on two-dimensional lattice for $q=2,3$, and $4$.
However, our present model on two-dimensional lattice undergoes a first-order phase transition with spontaneous $q$-fold symmetry breaking ($q=2,3$, and $4$) due to entropy effect of invisible states.
The model is fundamental for the analysis of a first-order phase transition with spontaneous discrete symmetry breaking.
}

\begin{document}

\maketitle

It has been a central issue in statistical physics to study the nature of phase transition with spontaneous breaking (SB) of symmetry since the establishment of the Ginzburg-Landau theory.
Recently, the concept of phase transition has not just been for natural phenomena but also for problems on information science, for example, coloring problem and the so-called satisfiability problem.
Study on phase transition with SB of symmetry has become increasingly important.
The Ising model and Potts model have been established as minimal models for the analysis of phase transition with SB of discrete symmetry.
The spin-$1/2$ ferromagnetic Ising model without an external magnetic field undergoes a second-order phase transition with SB of twofold symmetry at finite temperature.
Similarly, in the ferromagnetic $q$-state Potts model with no external magnetic field, $q$-fold symmetry is spontaneously broken at finite temperature.
It is important to investigate the relation between the order of the phase transition and the symmetry that breaks spontaneously at the transition point.

In the $q$-state Potts model, all spins take $q$ states and the interaction between spins is represented by Kronecker's delta~\cite{Potts-1952}.
Since the Potts model for $q=2$ is equivalent to the Ising spin systems, it is regarded as the straightforward extension of the Ising model.
In the ferromagnetic $q$-state Potts model on two-dimensional lattice, a second-order phase transition occurs at finite temperature when $q \le 4$, while a first-order phase transition occurs when $q > 4$~\cite{Wu-1982}.
The Potts model has been studied by analytical approaches as well as numerical methods~\cite{Potts-1952,Wu-1982,Baxter-1973,Kasteleyn-1969,Fortuin-1972,Binder-1981,Swendsen-1987,Ferrenberg-1988,Chen-1996,Feng-2008,Duff-2009,Kikuchi-1990,Suzuki-1967,Miyazima-1984,Murase-2008,Ohzeki-2006,Nishino-1998,Katori-1988,Kikuchi-1992,Miyashita-1979,Chang-2009,Hellmann-2009,Tomita-2001,Tomita-2002,Honmura-1984}.
To analyze the phase transition with SB of $q$-fold symmetry, it is enough to consider the $q$-state ferromagnetic Potts model in many circumstances.
The Potts model has succeeded in the analysis of phase transitions in experimental systems as well as theoretical systems~\cite{Wu-1982}.
Recently, a number of counterexamples have been found in some kinds of two-dimensional model~\cite{Tamura-2008,Stoudenmire-2009,Okumura-2010}.
Even if systems undergo a phase transition with SB of $q$-fold symmetry, the order of the phase transition of these systems is different from that of the ferromagnetic $q$-state Potts model.
To study these phenomena in a more systematic way, we introduce redundant states in the standard $q$-state Potts model. 
Now, we assume that redundant states do not contribute to the internal energy, while they affect the entropy and free energy.
In this paper, we focus on the consequence of introducing redundant states on the nature of the phase transition.

There are two main types of model where the order of the phase transition changes by controlling parameters.
Typical examples of the first type are the Blume-Capel model~\cite{Blume-1966,Capel-1966} and Blume-Emery-Griffiths (BEG) model~\cite{Blume-1971}.
In these models, the order of the phase transition changes when the parameter set such as a chemical potential is manipulated.
The other type is the model where the order of the phase transition varies if the number of microscopic states changes (e.g., the Wajnflasz model)~\cite{Wajnflasz-1971,Miyashita-2005,Boukheddaden-2005}.
The Wajnflasz model is regarded as the generalized Ising model in which the numbers of $+1$-states and $-1$-states are different.
In this model, the entropy-driven magnetic field is induced due to the bias of the numbers of $+1$-states and $-1$-states. 
As a result, the order of phase transition can be changed.
However, this phase transition does not accompany the breaking of $Z_2$ symmetry.
Although our present model is similar to the latter case,
the phase transition accompanies the breaking of $q$-fold symmetry.
Thus, our model is specialized for the analysis of the relation between the SB of symmetry and the order of phase transition.
Since our present model is the generalized Potts model as we discuss later, it can be applied to the analysis of SB of $q$-fold symmetry.

Our model is defined by the following Hamiltonian,
\begin{eqnarray}
 \label{Eq:Ham_qr}
  &&{\cal H} = -J \sum_{\left\langle i,j \right\rangle}
  \delta_{s_i,s_j}
  \sum_{\alpha=1}^q \delta_{s_i,\alpha}
  \delta_{s_j,\alpha}, 
  \\
 \label{Eq:condition_Ham_qr}
  &&s_i = 1,\cdots,q,q+1,\cdots,q+r,
\end{eqnarray}
where $\delta_{s_i,s_j}$ denotes Kronecker's delta. 
We consider the case of ferromagnetic interaction ($J>0$).
Hereafter, we call this model ($q$,$r$)-state Potts model, where $N$ is the number of spins.
The total number of microscopic states of this model is $(q+r)^N$, whereas that of the standard $q$-state Potts model is $q^N$.
We call redundant states, $q+1 \le s_i \le q+r$, ``invisible states''.
Following Eq.~(\ref{Eq:Ham_qr}), if and only if $1 \le s_i = s_j \le q$, the interaction is nonzero.
This Hamiltonian for $r=0$ is equivalent to the standard $q$-state Potts model.
Note that the internal energy does not change when the invisible states are introduced.
Thus, the ground state of this system has no invisible state and the number of ground state is $q$.
It is clear that a phase transition with SB of $q$-fold symmetry occurs in the present model. 
It is easy to show that the ($q$,$r$)-state Potts model is equivalent to the ($q$,1)-state Potts model with an external field depending on the temperature.
Introducing new spin variables, $\sigma_i=0,1,2,\cdots,q$ related to the original spins by $\sigma_i=s_i$ (if $s_i=1, \cdots, q$), $\sigma_i=0$ (otherwise) as the invisible state,
and taking the trace over $s_i$, we obtain the following Hamiltonian:
\begin{eqnarray}
 \label{Eq:Ham_Tlogr}
  {\cal H}' &=& -J \sum_{\left\langle i,j \right\rangle}
  \delta_{\sigma _i, \sigma _j}
  \sum_{\alpha=1}^q \delta_{\sigma _i,\alpha}
  \delta_{\sigma _j,\alpha}  -T \log r \sum_i \delta_{\sigma_i,0},
\end{eqnarray}
where $T$ represents temperature and we set the Boltzmann constant $k_{\rm B}$ to unity.
Note that the partition function of Eq.~(\ref{Eq:Ham_qr}) and that of Eq.~(\ref{Eq:Ham_Tlogr}) are the same.
The total number of microscopic states of this representation is $(q+1)^N$.
The second term in Eq.~(\ref{Eq:Ham_Tlogr}) comes from reducing the degrees of freedom of invisible states in the original Hamiltonian given by Eq.~(\ref{Eq:Ham_qr}), and it corresponds to the entropy effect.
This term represents the chemical potential for the invisible states, which induces the creation of invisible states at high temperature.

In this paper, we consider the order of the phase transition of the ferromagnetic ($q$,$r$)-state Potts model on a square lattice with periodic boundary condition.
We focus on the cases of $q=2,3,$ and $4$, because the standard ferromagnetic Potts model on two-dimensional lattice has a second-order phase transition for these values.

\begin{figure}[t]
  \begin{center}
   \includegraphics[width=0.36\columnwidth]{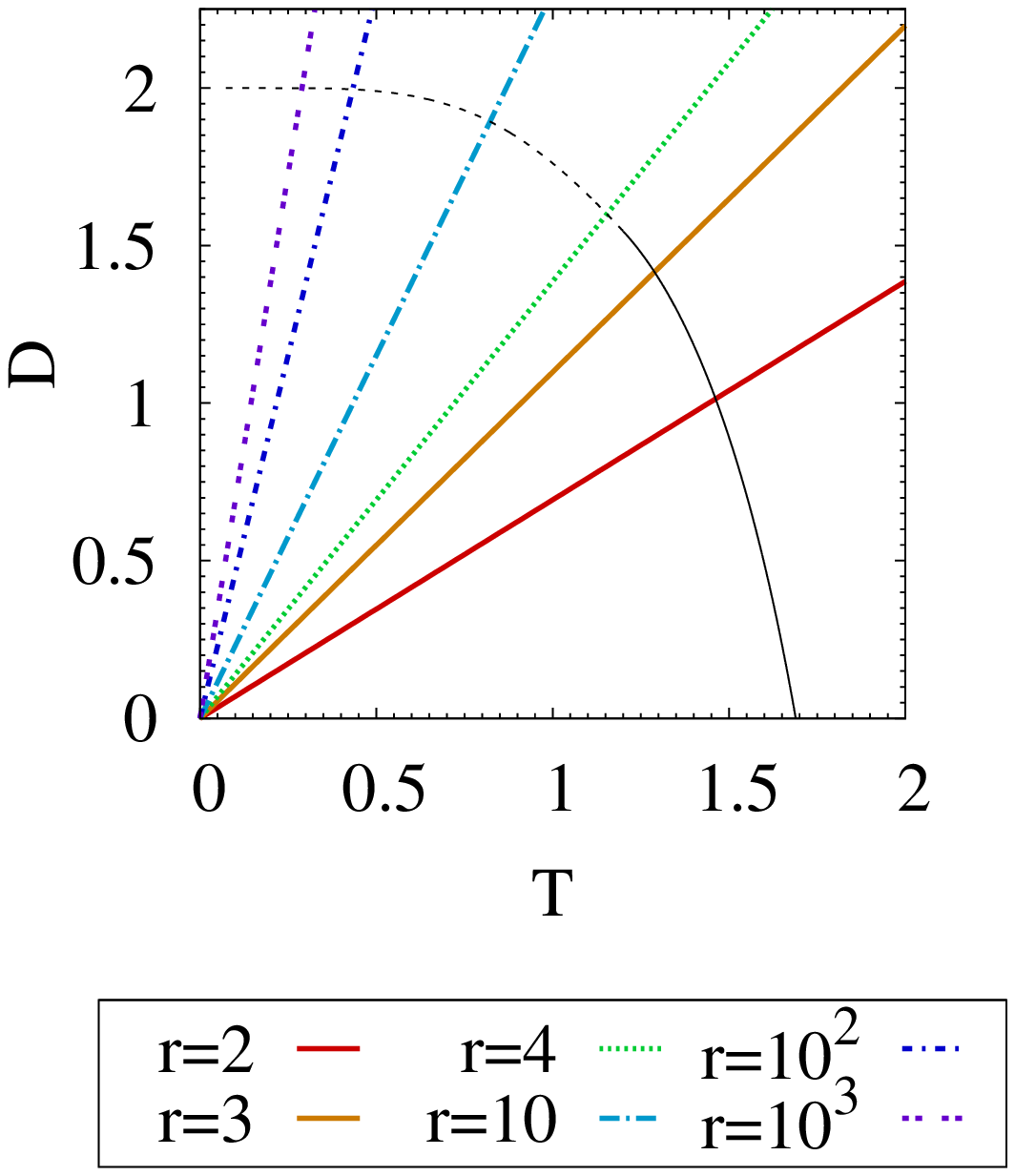}
   \includegraphics[width=0.56\columnwidth]{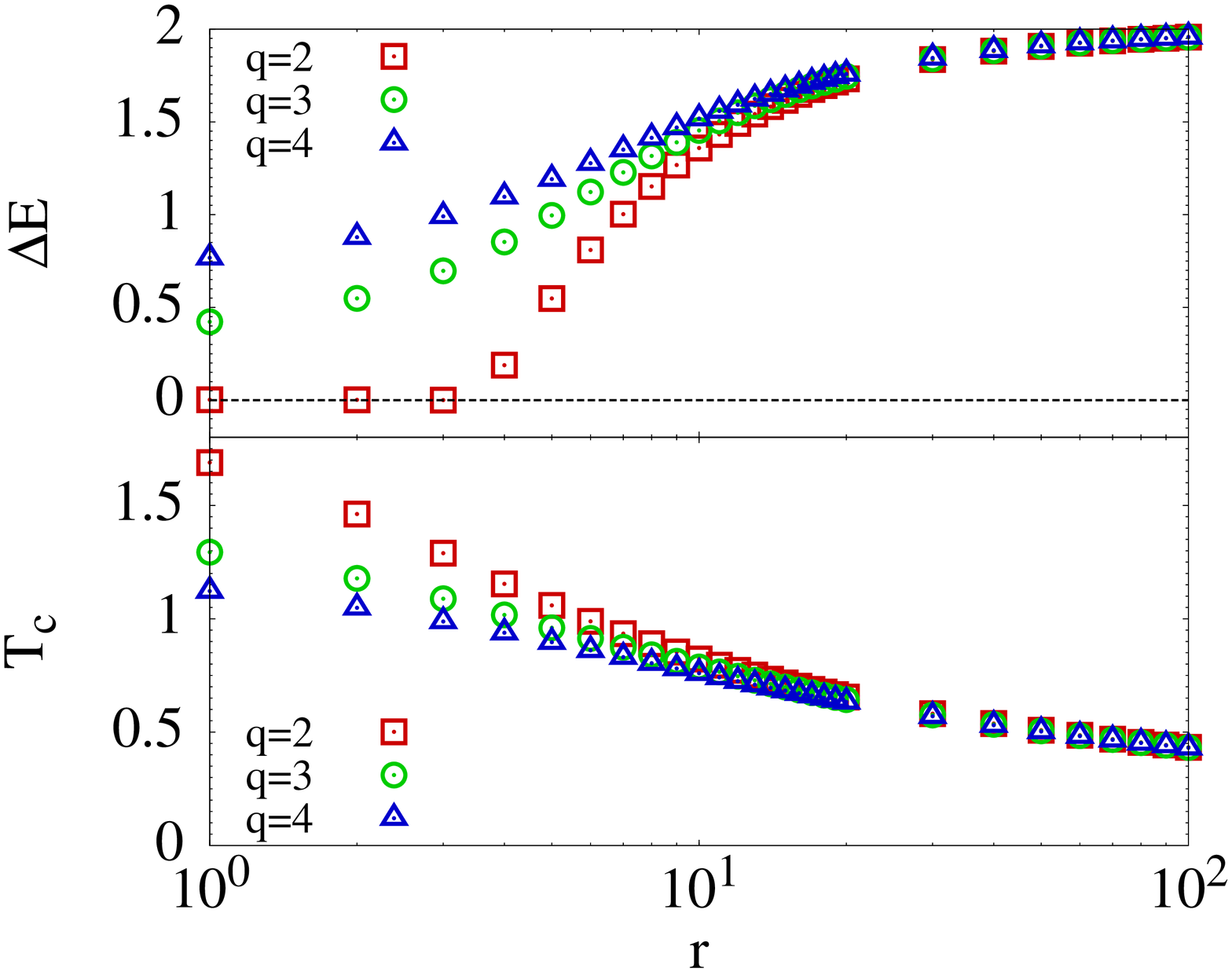}
   \end{center}
  \caption{
  \label{graph:mf-q2}
  (color online) Results of the mean-field calculation for the BEG model (Eq.~(\ref{Eq:Ham_beg_no_t_depend})).
 (left) Phase diagram.
  The dotted curve and solid curve represent first-order and second-order phase transition points, respectively.
  Tilted lines represent the trajectories of decreasing temperature for $r=2,3,4,10,10^2,$ and $10^3$ (from bottom to top).
  (right) Latent heat $\Delta E$ and transition temperature $T_{\rm c}$ as a function of $r$.
  }
 \end{figure}

Before considering the nature of the phase transition on a two-dimensional lattice, we analyze the phase transition of this model by mean-field calculation based on Bragg-Williams approximation with two parameters that represent the order parameter and density of invisible states.
We first consider the ($2$,$r$)-state Potts model that can be mapped on the BEG model~\cite{Blume-1971}:
\begin{eqnarray}
 \label{Eq:Ham_beg_no_t_depend}
 {\cal H}_{\rm BEG} &=& -\frac{J}{2} \sum_{\left\langle i,j \right\rangle}
 \left( t_i t_j + t_i^2 t_j^2 \right)
 - D \sum_i \left( 1-t_i^2 \right),\\
  t_i &=& +1, 0, -1,
\end{eqnarray}
where $D$ represents the crystal field. Hereafter, we take $J$ as the energy unit.
When $D = T \log r$, this Hamiltonian is equivalent to the ($2$,$r$)-state Potts model.
Now, we consider the case that the coordination number is four corresponding to a square lattice.
The left panel of Fig.~\ref{graph:mf-q2} shows the phase diagram of the BEG model obtained by mean-field calculation.
The dotted curve and solid curve indicate first-order and second-order phase transition points, respectively.
Thus, the decrease in temperature in the ($2$,$r$)-state Potts model corresponds to a tilting trajectory followed by $D=T\log r$ in the phase diagram shown in the left panel of Fig.~\ref{graph:mf-q2}.
As $r$ increases, the transition temperature decreases.
In the right panel of Fig.~\ref{graph:mf-q2}, we show the latent heat and transition temperature as functions of $r$ for $q=2,3,$ and $4$.

A second-order phase transition takes place only if ($q$,$r$) = ($2$,$1$), ($2$,$2$), and ($2$,$3$).
Note that the standard $q$-state ferromagnetic Potts model ($q \ge 3$ and $r=0$) has a first-order phase transition as far as mean-field calculation~\cite{Kihara-1954}.
As $r$ increases, the latent heat increases and the transition temperature decreases.
For $r \to \infty$, the latent heat and transition temperature approach $2$ and $0$, respectively.
From the mean-field calculation, we expect that the order of the phase transition changes by introducing invisible states.

\begin{figure}[t]
  \begin{center}
   \includegraphics[width=0.45\columnwidth]{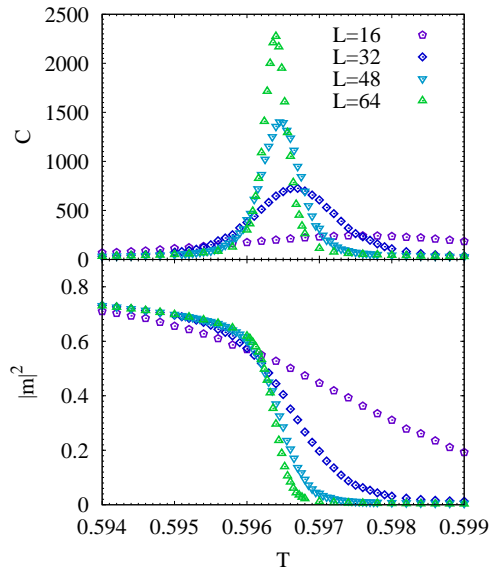}
  \end{center}
  \caption{
  \label{graph:eng_ord_vs_T}
  (color online)
Specific heat $C$ and the square of the order parameter $\left| \boldsymbol{m} \right|^2$ as functions of temperature of the ($3$,$25$)-state Potts model. The error bars are omitted for clarity since they are smaller than the symbol size.
  }
 \end{figure}

We also investigate the nature of the phase transition of the ($q$,$r$)-state Potts model by Monte Carlo simulation using the standard Metropolis method.
Each run contains $10^7-10^8$ Monte Carlo steps per spin at each temperature.
We make 8-16 independent runs for each size to evaluate the statistical errors.
Figure \ref{graph:eng_ord_vs_T} shows the specific heat $C$ and square of the order parameter of the ($3$,$25$)-state Potts model as a function of temperature.

It is convenient to introduce another representation of Kronecker's delta as $\delta_{\alpha, \beta} = \left[ 1 + \left( q - 1 \right) \boldsymbol{e}^\alpha \cdot \boldsymbol{e}^\beta \right]/q$, where $\boldsymbol{e}^\alpha$ ($\alpha = 1, 2, \cdots, q$) represents $q$ unit vectors pointing in the $q$ symmetric direction of a hypertetrahedron in $q-1$ dimensions~\cite{Wu-1982}.
The conventional order parameter for the $q$-state Potts model is given by $\boldsymbol{m} = \sum_{i=1}^N \boldsymbol{e}^{\sigma_i} / N$.
Note that in the ($q$,$r$)-state Potts model, we adopt the standard order parameter and the relation that $\boldsymbol{e}^{\sigma_i = 0} = 0$ for the invisible state.
As the system size $N(=L^2)$ increases, the maximum value of the specific heat increases linearly.
The square of the order parameter $\left| \boldsymbol{m} \right|^2$, which detects SB of $q$-fold symmetry, takes a finite value.
These facts indicate that a first-order phase transition with SB of $q$-fold symmetry occurs at finite temperature.

\begin{figure}[t]
  \begin{center}
   \includegraphics[width=0.5\columnwidth]{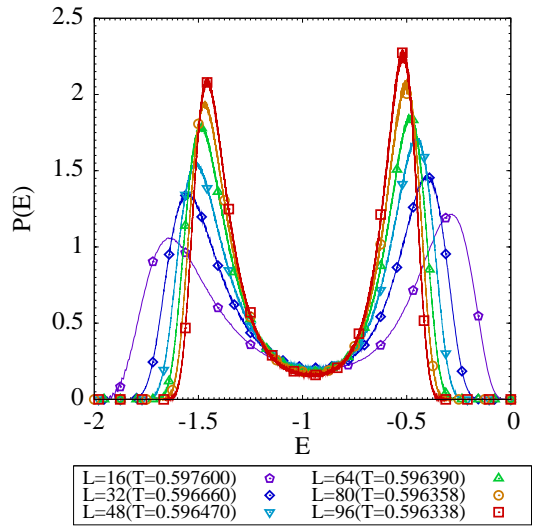}%
  \end{center}
  \caption{
  \label{graph:eng_histogram}
  (color online)
  Probability distribution of the internal energy of the ($3$,$25$)-state Potts model.
  The error bars are omitted for clarity since they are smaller than the symbol size.
  }
 \end{figure}

\begin{figure}[t]
  \begin{center}
   \includegraphics[width=0.5\linewidth]{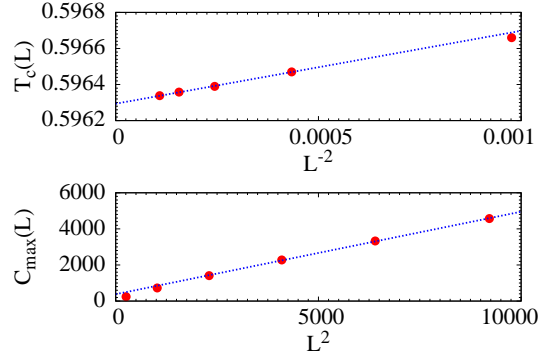}%
   \end{center}
  \caption{
  \label{graph:estimate}
  (color online) Finite-size scaling for the ($3$,$25$)-state Potts model.
 (Top panel) Transition temperature depending on the lattice size as a function of $L^{-2}$.
  (Bottom panel) Maximum value of the specific heat as a function of $L^2$.
Dotted lines represent fitting curves using data of $L \ge 48$.
The error bars are omitted for clarity since they are smaller than the symbol size.
  }
 \end{figure}

To obtain the transition temperature and latent heat, we calculate the probability distribution of the internal energy and apply finite-size scaling for a first-order phase transition following the method in Ref.~36).
Figure \ref{graph:eng_histogram} shows the probability distribution of the internal energy $P\left( E\right)$, which has a form of bimodal distribution.
The bimodal distribution of energy histogram is considered to be a typical behavior of a first-order phase transition.
The maximum values of the size-dependent specific heat $C_{\rm max} (L)$ and the transition temperature $T_{\rm c} (L)$ depending on the lattice size defined by $C(T_{\rm c}(L)) = C_{\rm max} (L)$ are obtained by the reweighting method.
We adopt the finite-size scaling relations for a first-order phase transition in a $d$-dimensional system~\cite{Challa-1986}:
\begin{eqnarray}
 \label{Eq:estimate_tc}
  T_{\rm c} (L) &=& aL^{-d} + T_{\rm c}, \\
 \label{Eq:estimate_lh}
  C_{\rm max} (L) &\propto& \frac{(\Delta E)^2 L^d}{4T_{\rm c}^2},
\end{eqnarray}
where $\Delta E$ and $T_{\rm c}$ represent the latent heat and transition temperature for infinite systems, respectively.
The top panel of Fig.~\ref{graph:estimate} shows $T_{\rm c} (L)$ as a function of $L^{-2}$.
We fit the data using Eq.~(\ref{Eq:estimate_tc}) and obtain the transition temperature $T_{\rm c} = 0.59630(1)$.
This value is lower than that of the standard 3-state Potts model~\cite{Wu-1982}.
The bottom panel of Fig.~\ref{graph:estimate} shows $C_{\rm max} (L)$ as a function of $L^2$.
We fit the data using Eq.~(\ref{Eq:estimate_lh}) and we obtain $\Delta E = 0.81(2)$ using the transition temperature given by the above calculation.
Using the same technique, we obtain the transition temperature and latent heat of ($2$,$30$)- and ($4$,$20$)-state Potts models which are shown in Table \ref{table:tc_lh}.
We confirmed a first-order phase transition for these cases as well.

\begin{table}
\begin{center}
 \caption{\label{table:tc_lh}
Transition temperature and latent heat for ($2$,$30$)-, ($3$,$25$)-, and ($4$,$20$)-state Potts models.}
\begin{tabular}{c|ccc}
\hline\hline
 ($q$,$r$) & ($2$,$30$) & ($3$,$25$) & ($4$,$20$) \\
 \hline
 $T_{\rm c}$ & $0.57837(1)$ & $0.59630(1)$ & $0.61683(1)$ \\
 $\Delta E$ & $1.02(2)$ & $0.81(2)$ & $0.68(2)$\\
\hline
\end{tabular} 
\end{center}
\end{table}

The ($q$,$r$)-state Potts model is expected to be a fundamental model for analyzing a first-order phase transition with SB of $q$-fold ($q=2,3$, and $4$) symmetry in a two-dimensional lattice.
It is found that a first-order phase transition occurs at finite temperature on two-dimensional fully frustrated continuous spin systems with SB of threefold symmetry~\cite{Tamura-2008,Stoudenmire-2009,Okumura-2010}.
Since these models are based on continuous spin systems, these have additional degrees of freedom in addition to three states.
Thus, it is possible that these degrees of freedom can be regarded as the existence of intrinsic invisible states.
Furthermore, since the Potts model is applied to a broad range of science, e.g., information science~\cite{Mulet-2002,Kurihara-2009,Sato-2009}, we also expect that the ($q$,$r$)-state Potts model is widely applicable for information science and technology.

To summarize, we have studied a finite-temperature phase transition in the ferromagnetic Potts model with invisible states on two-dimensional square lattice.
The invisible states are introduced as redundant states that affect the entropy and free energy.
However, they do not contribute to the internal energy.
We found that a thermal-driven first-order phase transition occurs in this system for $q=$ 2, 3, and 4 due to the entropy effect of invisible states.
We also found that the latent heat increases with the number of invisible states $r$ in the mean-field calculation as stated above.
Thus, we expect that a first-order phase transition occurs for larger $r$ than those used in the Monte Carlo simulation.
To find a boundary of $r$ between occurring a second-order and a first-order phase transition is an issue in the future.
The present model can be regarded as the generalized Potts model that is fundamental for analysis of spontaneous discrete symmetry breaking.
We believe that the present model provides novel insight into such kind of phase transition.

The authors are grateful to Jie Lou, Yoshiki Matsuda, Seiji Miyashita, Takashi Mori, Yohsuke Murase, Masayuki Ohzeki, and Eric Vincent for their valuable comments.
R.T. is partially supported by Global COE Program ``the Physical Sciences Frontier'', MEXT, Japan.
S.T. is partially supported by Grant-in-Aid for Young Scientists Start-up (21840021) from the JSPS.
The present work is financially supported by MEXT Grant-in-Aid for Scientific Research (B) (19340109,22340111), and for Scientific Research on Priority Areas ``Novel States of Matter Induced by Frustration'' (19052004), and by Next Generation Supercomputing Project, Nanoscience Program, MEXT, Japan.
The computation in the present work was performed on computers at the Supercomputer Center, Institute for Solid State Physics, University of Tokyo. 


\end{document}